\documentclass[final]{raa06}           
\usepackage{graphicx,times}             
\usepackage{natbib}
\usepackage{amssymb,amsmath}
\bibpunct{(}{)}{;}{a}{}{,}
\usepackage[colorlinks=true, citecolor=blue]{hyperref}%

\usepackage{graphicx,kantlipsum,setspace}
\usepackage{caption}
\usepackage{graphics,epsf}
\usepackage{amsmath}                
\usepackage{amsfonts}               
\usepackage{amssymb}                
\usepackage{epsfig}                 
\usepackage{appendix}
\usepackage{graphicx}
\usepackage{float}
\usepackage{color}
\usepackage{multirow}
\usepackage{colortbl}
\usepackage[para,online,flushleft]{threeparttable}
\usepackage{xcolor}

\hypersetup{citecolor=blue, 
            linkcolor=red, 
            menucolor=blue, 
            urlcolor=blue}  

\newcommand{\kms}{{~\rm km\; s^{-1}}}

\newcommand{\msyr}{{~M_{\odot}~\rm yr^{-1}}}
\newcommand{\cm}{{~\rm cm}}
\newcommand{\m}{{~\rm m}}
\newcommand{\km}{{~\rm km}}
\newcommand{\s}{{~\rm s}}
\newcommand{\h}{{~\rm h}}

\newcommand{\g}{{~\rm g}}
\newcommand{\G}{{~\rm G}}
\newcommand{\K}{{~\rm K}}
\newcommand{\erg}{{~\rm erg}}
\newcommand{\yr}{{~\rm yr}}
\newcommand{\Myr}{{~\rm Myr}}
\newcommand{\Gyr}{{~\rm Gyr}}
\newcommand{\pc}{{~\rm pc}}
\newcommand{\kpc}{{~\rm kpc}}

\newcommand{\eV}{{~\rm eV}}
\newcommand{\keV}{{~\rm keV}}
\newcommand{\kev}{{~\rm keV}}
\newcommand{\AU}{{~\rm AU}}
\newcommand{\mum}{{~\rm \mu m}}

\def \astrobj#1{#1}

\begin{document}

   \title{The morphology of supernova remnant G0.9+0.1 implies explosion by jittering-jets
}

   \volnopage{Vol.0 (20xx) No.0, 000--000}      
   \setcounter{page}{1}          


   \author{Noam Soker
    }

   \institute{Department of Physics, Technion - Israel Institute of Technology, Haifa, 3200003, Israel;   {\it     soker@physics.technion.ac.il}\\
\vs\no
   {\small Received~~20xx month day; accepted~~20xx~~month day}}

\abstract{
I examine the morphology of the core-collapse supernova (CCSN) remnant (SNR) G0.9+0.1 and reveal a point-symmetrical morphology that implies shaping by three or more pairs of jets, as expected in the jittering jets explosion mechanism (JJEM). The large northwest protrusion, the ear (or lobe), has two bright rims. I compare this ear with its rims to an ear with three rims of a jet-shaped planetary nebula and jets from an active galactic nucleus that shaped several rims on one side. Based on this similarity, I argue that two jets or more shaped the northwest ear of SNR G0.9+0.1 and its two rims. I identified the bright region south of the main shell of SNR G0.9+0.1 as a jet-shaped blowout formed by a jet that broke out from the main SNR shell. I base this on the similarity of the blowout of SNR~G0.9+0.1 with that of SNR~G309.2-00.6, argued in the past to be shaped by jets. I identify four symmetry axes along different directions that compose the point-symmetric morphology of SNR G0.9+0.1. I show that the morphological features of holes, granular texture, and random filaments exist in CCSNe and planetary nebulae and are unlikely to result from some unique processes in CCSNe. These structures result from similar instabilities in the JJEM and the neutrino-driven explosion mechanism and, unlike a point-symmetric morphology, cannot determine the explosion mechanism. Identifying SNR G0.9+0.1 as a new point-symmetric CCSN strengthens the JJEM as the primary explosion mechanism of CCSNe.  
}

\maketitle

\section{Introduction}
\label{sec:Introduction}

The community of core-collapse supernovae (CCSNe) is far from a consensus on the explosion mechanism. The debate is mainly around the delayed neutrino explosion mechanism (neutrino-driven mechanism) and the jittering jets explosion mechanism (JJEM), two alternative theoretical explosion mechanisms to explode CCSNe. 

Recent studies of the neutrino-driven mechanism focused mainly on simulating the revival of the stalled shock by neutrino-heating of the collapsing core material and the consequences (e.g., \citealt{Andresenetal2024, BoccioliFragione2024, Burrowsetal2024kick, JankaKresse2024, vanBaaletal2024, WangBurrows2024, Laplaceetal2025, Huangetal2025, Bocciolietal2025, EggenbergerAndersenetal2025, Imashevaetal2025, Maltsevetal2025, Maunderetal2025, Mulleretal2025, Nakamuraetal2025, SykesMuller2025, Janka2025, ParadisoCoughlin2025, WangBurrows2025}, for some papers from the last two years). 
The magnetorotational explosion mechanism, where one pair of jets with a fixed axis explode the star (e.g., \citealt{Kondratyevetal2024, Shibagakietal2024, ZhaMullerPowell2024, Shankaretal2025, Shibataetal2025}), occurs in rare ($\simeq 1\%$ of CCSNe; e.g., \citealt{Muller2024}) cases when the pre-collapse core is rapidly rotating. I group the magnetorotational explosion mechanism with the neutrino-driven mechanism because the magnetorotational explosion mechanism attributes most CCSNe to the neutrino-driven mechanism.

The JJEM assumes that the newly born neutron star (NS) launches $\simeq 5$ to $\simeq 30$ pairs of opposite jets (e.g., \citealt{Soker2024Learning}) that explode the star, even in electron capture CCSNe \citep{WangShishkinSoker2024}. Neutrino heating adds energy to the jets but is secondary to the jets \citep{Soker2022nu}. The jets' power and direction vary stochastically, leading to point-symmetric morphologies (e.g., \cite{Braudoetal2025}). In the JJEM, the accretion of material with stochastic angular momentum onto the newly born NS forms intermittent accretion disks with varying angular momentum axes (e.g., \citealt{ShishkinSoker2021, ShishkinSoker2022}); these intermittent accretion disks launch the jittering jets. 

Some of the outcomes of the neutrino-driven mechanism and the JJEM are similar, while others differ \citep{Soker2024Univ2}. 
Similar, but not identical, are the overall nucleosynthesis and neutrino emission processes. Gravitational wave emission, as one example, differs between the neutrino mechanism (e.g., \citealt{Mezzacappaetal2023}) and the JJEM (\citealt{Soker2023GW}), but gravitational wave observations are far from the ability to distinguish between the two mechanisms. One challenge of the neutrino-driven mechanism is to explain CCSNe with an explosion energy of $\gtrsim 2 \times 10^{51} \erg$, even when the energy of magnetars is included (e.g., \cite{KumarA2025}), and in some gamma ray bursts (e.g., \citealt{Fioreetal2025}).  

Supernova remnants (SNRs) can teach us a lot about the physics of the interaction of the ejecta with the ambient gas, including cosmic ray acceleration, emission properties (e.g., \citealt{Yamazakietal2014RAA, Zhangetal2016RAA, Lietal2020RAA, Luoetal2024RAA}), interaction of the ejecta with the ambient gas (e.g., \citealt{Yanetal2020RAA, Luetal2021RAA}) including jets (e.g., \citealt{YuFang2018RAA}), the role of the NS remnant in the center (e.g., \citealt{HorvathAllen2011RAA, Wuetal2021RAA}), magnetohydrodynamics (e.g., \citealt{Wuetal2019RAA, Leietal2024RAA}), and, most relevant to this study, about the morphology of the ejecta (e.g., \citealt{Renetal2018RAA}). I will use morphological considerations to prefer the JJEM over the neutrino-driven mechanism. 

The best properties to indicate the explosion mechanism of the majority, and likely all, CCSNe, are the presence or absence of signatures of jet activity in the explosion morphology. Since the morphologies of CCSNe are unresolved, the most robust approach is to study the CCSN remnant (CCSNR) morphologies. Most prominent is the prediction of the JJEM that in many, but not all, CCSNRs, two or more pairs of jittering jets will imprint opposite (to the center) structural features. Namely, the jittering jets might shape a point-symmetric CCSNR; this most clearly appears in the three-dimensional hydrodynamical simulations of the JJEM by \cite{Braudoetal2025}. 
The neutrino mechanism does not explain most point-symmetric CCSNRs' properties (\citealt{SokerShishkin2025Vela}). 
Indeed, \cite{Vartanyanetal2025} presented a simulation of the neutrino-driven mechanism until the shock breaks out from the star but did not reproduce a point-symmetric morphology (and ignored this possibility). 
Earlier studies showed the difficulties of instabilities in the neutrino-driven mechanism to account for the morphology of the jet in Cassiopeia A (e.g., \citealt{Soker2017RAACas87A}) and for the iron structure of SN 1987A  (e.g., \citealt{Soker2017RAACas87A, BearSoker2018}).

By systematically identifying point-symmetric morphologies in over ten CCSNRs and connecting these morphologies to shaping by jets in the frame of the JJEM, researchers have established the JJEM as the primary explosion mechanism of CCSNe (for a review, see \citealt{Soker2024Rev}). After the identification of a point-symmetric structure in the Doppler map of supernova remnant (SNR) 0540-69.3 \citep{Soker2022SNR0540}, came a list of CCSNRs with identified jet-shaped point-symmetric morphologies: 
CTB~1 \citep{BearSoker2023RNAAS}, 
the Vela CCSNR (\citealt{Soker2023SNRclass, SokerShishkin2025Vela}), 
the Cygnus Loop \citep{ShishkinKayeSoker2024},
N63A \citep{Soker2024CounterJet},
SN 1987A \citep{Soker2024NA1987A, Soker2024Keyhole}, 
G321.3–3.9 \citep{Soker2024CF, ShishkinSoker2025G321},
G107.7-5.1 \citep{Soker2024CF},
W44 \citep{Soker2024W44}, 
Cassiopeia A \citep{BearSoker2025}, 
Puppis A \citep{Bearetal2025Puppis},  
the Crab Nebula \citep{ShishkinSoker2025Crab},
S147 \cite{Shishkinetal2025S147},
and N132D (\citealt{Soker2025N132D}).

In Section \ref{sec:SNRG0901}, I identify point-symmetric morphological features in the SNR G0.9+0.1 and rims along a lobe that suggest more than two jet-launching episodes, compatible with the JJEM.   

\cite{Orlandoetal2025Holes} simulated the formation of holes and \cite{Orlandoetal2025Filaments} of filaments in Cassiopeia A in the frame of the neutrino mechanism. In Section \ref{sec:Instabilities}, I claim that the filaments cannot distinguish between the two explosion mechanisms, and therefore, the simulations by \cite{Orlandoetal2025Holes} and \cite{Orlandoetal2025Filaments} cannot serve to argue for or against the two explosion mechanisms. 
In Section \ref{sec:Summary}, I summarize this study that adds to the accumulating evidence for the JJEM. 

\section{Jet-shaped morphological features in SNR G0.9+0.1}
\label{sec:SNRG0901}

I am going to identify morphological features by eye inspection. 
Although this method might seem completely subjective, it is not. The eye inspection method to identify pairs of symmetrical structural features has a significant objective component. Humans possess a remarkable ability to critically examine mirror symmetry and identify departures from it by inspection alone. This is because facial symmetry is a key indicator of mate quality (e.g., \citealt{Rhodes2006, Pinheiroetal2023} and more references therein). Particularly, humans' ability to identify \textit{fluctuating asymmetries}, which are random (non-directional) deviations from perfect symmetry in bilaterally paired traits, is crucial because fluctuating asymmetries tend to reflect health problems (e.g., \citealt{Rhodes2006}). Most astrophysicists, accustomed to quantitative calculations, have made one of the most important decisions of their lives, if not the most important, i.e., choosing a mate, primarily through visual inspection (as the present author did), particularly by looking for symmetry and ruling out large fluctuating asymmetries. This method is therefore objective in large part because recognition of symmetry is relatively automatic and consistent across cultures (e.g., \citealt{Rhodes2006}).

\subsection{The general morphology of SNR G0.9+0.1}
\label{subsec:GeneralMorphology}
\begin{figure*}
\begin{center}
\includegraphics[trim=0.0cm 8.5cm 0.0cm 0.0cm,width=0.98\textwidth]{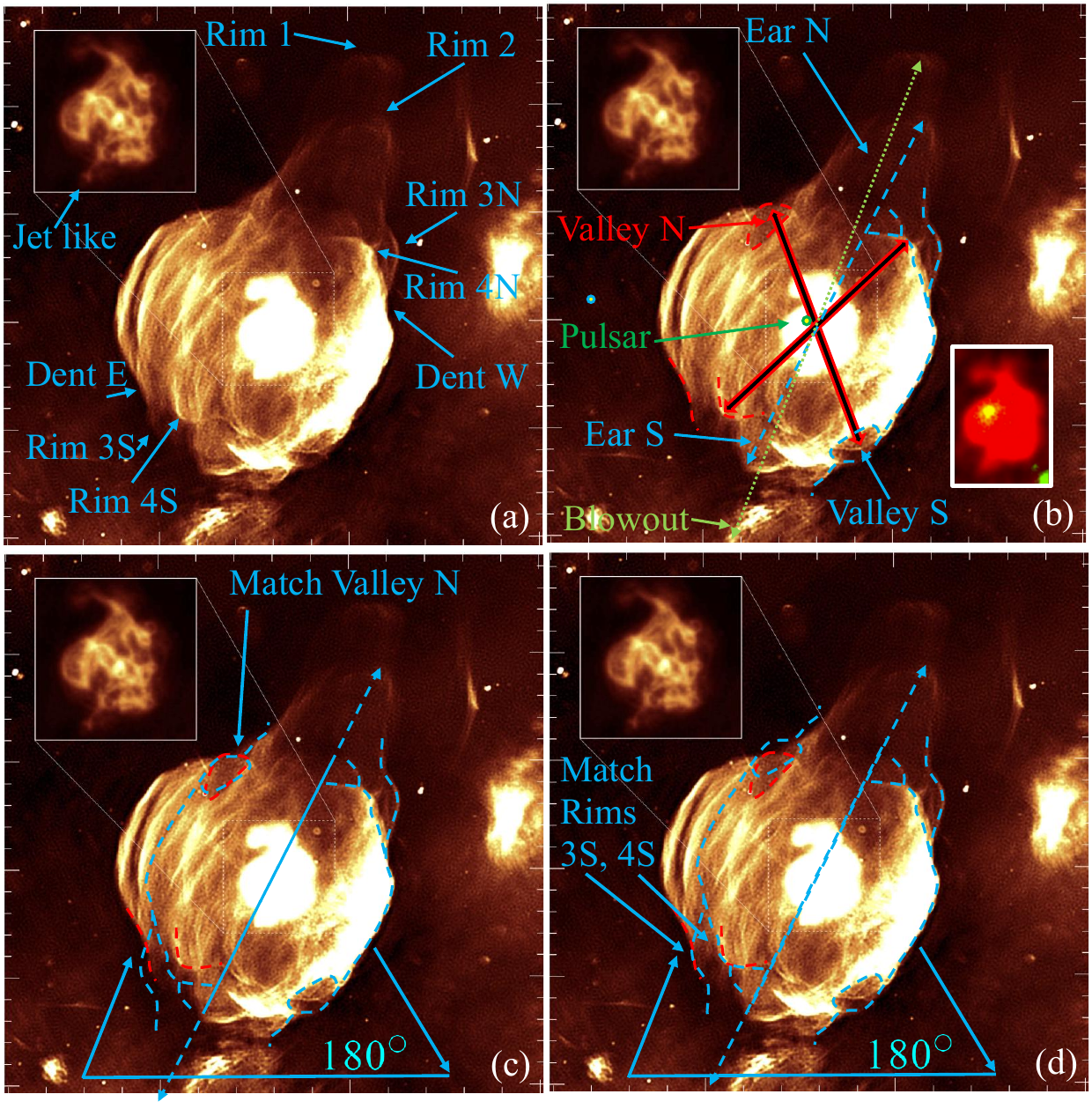} 
\caption{A MeerKat radio image at 1.28 GHz of SNR~G0.9+0.1 adapted from \cite{MeerKAT2022}. The inset on the upper left is a desaturated image of the PWN. I added all the marks on the different panels. (a) My marks of some relevant structural features for this study. (b) The dashed pale blue and the green-dotted two-sided arrows mark the axes of two pairs of jets that this study proposes to compose the main-jet axis of SNR~G0.9+0.1. The two-sided double-lined red arrows mark my identification of two additional symmetry axes. The lower right inset is an image from \cite{Camiloetal2009}, who discovered the pulsar in the yellow dot, a radio and X-ray bright region. I marked this location on the main image with a yellow-green dot, just to the northeast of the cross point of the two-sided arrows. The cross point is the center of the explosion. The two double-lined red arrows have their center at the cross point. I marked several morphological features and boundaries west of the main-jet axis with curved, dashed, pale blue lines. The dashed red lines mark some features east of the main jet axis. The four symmetry axes (double-sided arrows) form the `wind-rose' of the point-symmetric morphology. (c)  I rotated the western dashed pale blue lines structure by $180^\circ$ around the cross point of the double-sided arrows, and displaced it to the southeast to matched Valley S to Valley N; it matches some other morphological features on the east. (d) I rotated the western dashed pale blue lines structure by $180^\circ$ around the cross point of the double-sided lines. This matches Rim 3N and Rim 4N (dashed pale-blue lines) with Rim 3S and Rim 4S (dashed-red lines); see Section \ref{subsec:point}.    
}
\label{fig:SNRG0901}
\end{center}
\end{figure*}

SNR~G0.9+0.1 near the galactic center has a central zone, which is a pulsar wind nebula (PWN), surrounded by a more extended shell (e.g., \citealt{HelfandBecker1987}). \cite{GaenslerPivovaroffGarmire2001} studied a PWN in the inner region of SNR G0.9+0.1, and \cite{Camiloetal2009} discovered the pulsar. The main image in Figure \ref{fig:SNRG0901}, adapted from \cite{MeerKAT2022}, shows a radio image of SNR~G0.9+0.1, where the PWN is saturated; the inset shows the desaturated PWN. In this study, I refer only to the morphologies of objects and, hence, do not specify their angular size and other properties found in the papers I cite. 

\cite{MeerKAT2022} attribute the northwest protrusions, which I refer to as Ear N, and the rims to the activity of the pulsar in the center, particularly the jets it launches to the north (\citealt{Dubneretal2008}). 
I note the following. (1) The jet-like feature in the south of the PWN that \cite{GaenslerPivovaroffGarmire2001} identified is not the jets I refer to in this study. The jet-like feature does not expand to the outer regions of the SNR. (2) The present northern pulsar jet bends close to the center (bends to the east). (3) The energy to inflate the northern ear with its two rims, the two other pairs of rims (numbers 3 and 4), and the southern ear (Ear S), is tens of percents of the kinetic energy in the shell (estimated by the volume they occupy, as done by \cite{GrichenerSoker2017}). I suggest, therefore,  that the pulsar wind nebula fills the empty volume in the center rather than shapes the outer remnant. I attribute the main shaping of SNR~G0.9+0.1 to jets during the explosion process. The pulsar's rotation that now powers the PWN is a remnant of the accretion of angular momentum via intermittent accretion disks that launched the jittering jets during the explosion. In Section \ref{subsec:NorthernEar} I compare the northwest lobe to similar structures in two astrophysical objects, and in Section \ref{subsec:point} I identify a point symmetrical morphology that I attribute to jittering jets that are part of more jets that exploded the star.     

\subsection{The northwest ear}
\label{subsec:NorthernEar}

I refer to Ear N to the northwest and its Rim 1 and Rim 2. A double-rim structure in an ear or a lobe is a strong indication of multiple jet activity, as evident in planetary nebulae and from active galactic nucleus jets. 
Hydrodynamical simulations demonstrate the formation of rims, similar to those observed in planetary nebulae, by jets in conditions appropriate to planetary nebulae (e.g., \citealt{Balicketal2017, Balicketal2018, Balicketal2019}). Jets that explode the star in the frame of the JJEM also form such rims as \cite{Braudoetal2025} have demonstrated in a recent set of three-dimensional simulations of the JJEM.

\begin{figure}
\begin{center}
\includegraphics[trim=0cm 23.0cm 5.5cm 0cm,scale=0.62]{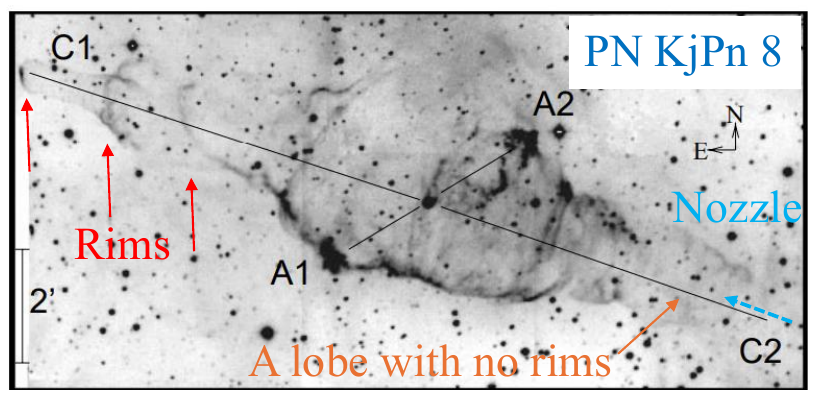} 
\caption{An image of the planetary nebula KjPn~8 adapted from \cite{Lopezetal2000}, who attributed the two axes, C1-C2 and A1-A2 to two pairs of jets. The black marks are from the original image. I mark the three rims on the eastern lobe. No such bright rims exist in the western lobe.  
}
\label{Fig:KjPn8}
\end{center}
\end{figure}

\begin{figure}
\begin{center}
\includegraphics[trim=0cm 18.2cm 3.5cm 0cm,scale=0.53]{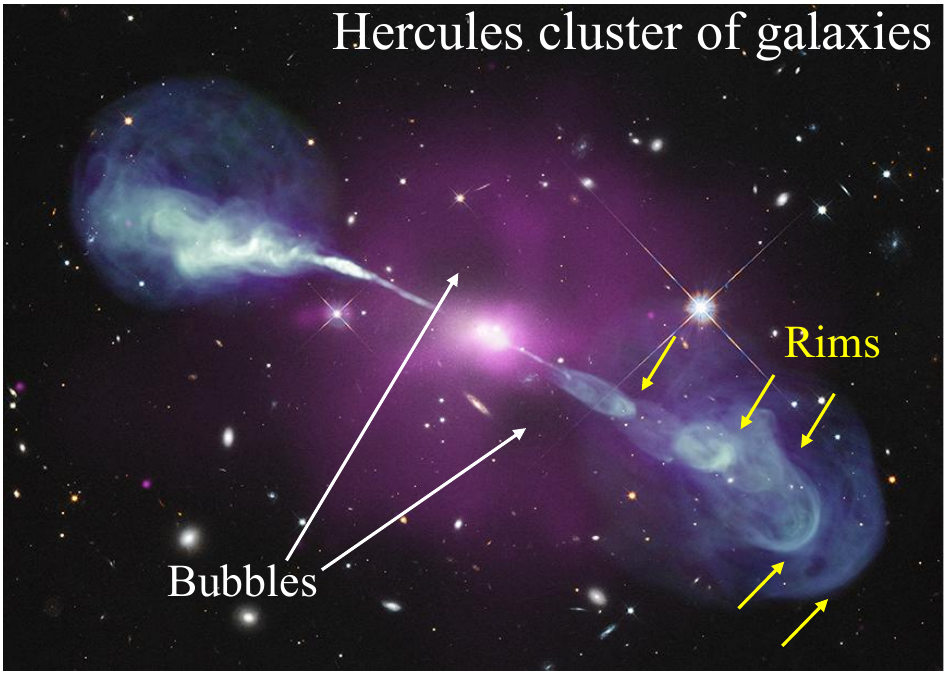} 
\caption{An image of the Hercules cluster of galaxies adapted from Chandra site (Credit: X-ray: NASA/CXC/SAO, Optical: NASA/STScI, Radio: NSF/NRAO/VLA;  \tiny\url{https://chandra.harvard.edu/photo/2014/archives/archives_herca.jpg}\normalsize). 
The radio emission (blue) reveals pairs of opposite jets, with several sharp rims formed by jets on the right side in the image. The X-ray emission (purple) reveals pairs of bubbles inclined to the radio jets (see also, e.g.,  \citealt{Ubertosietal2025}), namely, a point-symmetric morphology. Red and mainly white show the visible emission.
}
\label{Fig:Hercules}
\end{center}
\end{figure}

In Figure \ref{Fig:KjPn8}, I present an image of the point symmetric planetary nebula KjPn~8 adapted from \cite{Lopezetal2000}. Lobe C1 has three rims, while the counter lobe C2 has none. Namely, a pair of unequal lobes. \cite{Lopezetal2000} suggested that jets shaped the two C1 and C2 lobes, and the pair of ears A1 and A2. The two symmetry axes (black lines) are from their original figure. In the case of KjPn~8, the jets are not active anymore. In Figure \ref{Fig:Hercules}, multiple rim structures appear on one side of the Hercules cluster of galaxies (right side of the figure). 
The jets are still active in this case, and the two sides are unequal.  
The SNR candidate G107.7-5.1 has one lobe with three rims and an unequal opposite lobe (Figures 31-34 in the discovery paper by \citealt{Fesenetal2024}). In \cite{Soker2024CF}, I argue that jets shaped the morphology of the SNR candidate G107.7-5.1, applying similar arguments to the above arguments.         

\begin{figure*}[t]
\begin{center}
\includegraphics[trim=0.0cm 18.0cm 0.0cm 2.9cm,scale=0.65]{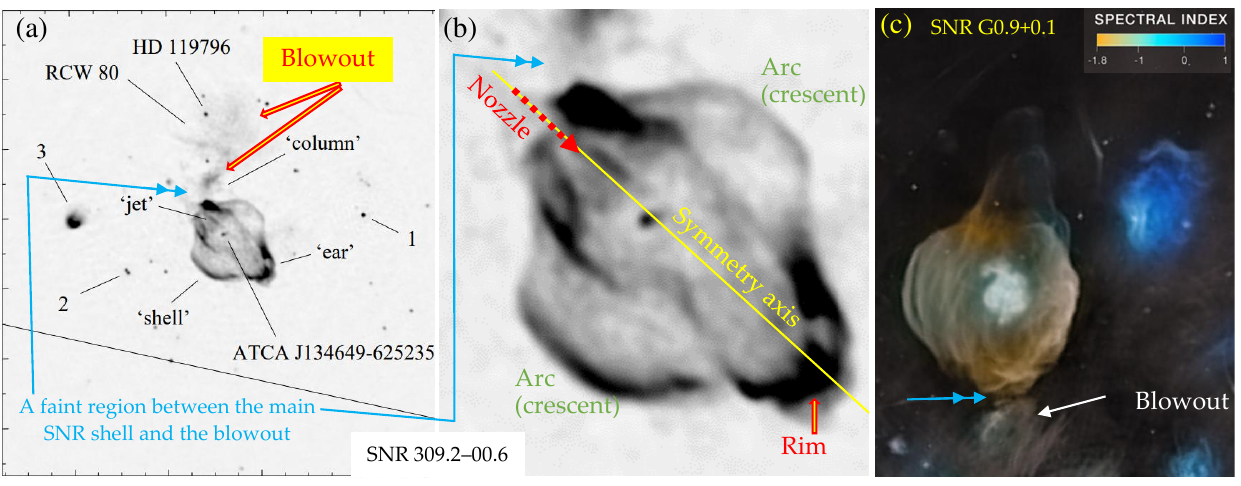} 
\caption{(a) A radio image of SNR~G309.2–00.6 adapted from \cite{Gaensleretal1998} with their marks in black; they already claimed that jets shaped this CCSN. \cite{ShishkinKayeSoker2024} added the mark of the blowout. I mark the faint zone between the main SNR shell and the blowout with double-headed arrows in this and the other panels. (b) The main CCSNR shell. \cite{Soker2024PNSN} marked the symmetry axis, the two arcs (`crescents’), and the rim-nozzle structure. (c) Radio spectral index image of SNR~G0.9+0.1 adapted from JWST site (Credit: NASA, ESA, CSA, STScI, SARAO, Samuel Crowe (UVA), John Bally (CU), Ruben Fedriani (IAA-CSIC), Ian Heywood (Oxford); 
\tiny\url{https://webbtelescope.org/contents/media/images/2025/115/01JQC49FHDYKY5WV5BH8BXSZFJ}\normalsize). 
The spectral index ranges from $-1.8$ in orange to 1 in pale blue. I added the labeling of the blowout and the faint zone between the blowout and the main SNR shell. 
 }
\label{fig:SNRG309}
\end{center}
\end{figure*}

Ear S on the southeast (panel b of Figure \ref{fig:SNRG0901}) is much smaller than Ear N on the other side of the SNR. I suggest that the counter structure to Ear N also includes the blowout, which I marked on panel b of Figure \ref{fig:SNRG0901} and on panel c of Figure \ref{fig:SNRG309}. There is a large blowout to the north of the  SNR~G309.2-0.6 and to the south of the SNR the Cygnus Loop that \cite{ShishkinKayeSoker2024} attributed to jets that brook out from the main SNR ejecta. They further argued that these jets were part of the exploding jets. 
Panels a and b of Figure \ref{fig:SNRG309} present SNR~G309.2-0.6 with its blowout. 
Based on these two SNRs and the analysis by \cite{ShishkinKayeSoker2024}, I term the bright region south of the main shell of SNR~G0.9+0.1 a blowout, and attribute it to a jet that broke out from the main shell, the counter jet to the one that shaped Rim 1 (dotted-green line on panel b of Figure \ref{fig:SNRG0901}). Panel c of Figure \ref{fig:SNRG309} shows that the radio spectral index of the blowout of SNR~G0.9+0.1 is similar to that of the main shell, and both are very different from another bright region to the west-northwest of the SNR. Moreover, in both SNR~G309.2-0.6 and SNR~G0.9+0.1, there is a faint region separating the main shell and the blowout, as I mark by double-headed pale-blue arrows on the three panels of Figure \ref{fig:SNRG309}.  

The blowout and Ear N define the main jet axis of SNR~G0.9+0.1 to be between the dashed-pale-blue and dotted-green lines on panel c of Figure \ref{fig:SNRG0901}. Rim 1 and Rim 2 of Ear N on one side and Ear S and the blowout on the other form a point-symmetric structure. To solidify the classification of SNR~G0.9+0.1 as a point-symmetric SNR, I turn to identify more point-symmetric morphological features.  

\subsection{Identifying a point-symmetric morphology}
\label{subsec:point}

As with over ten previously identified point-symmetrical CCSNRs (see list in Section \ref{sec:Introduction}), the observed jet-shaped morphology during the CCSNR phase is imperfect and not always prominent. The reason is that at least one, and more likely several, processes act to distort and smear the point symmetric morphology, including possible interaction with a circumstellar material and the interstellar medium, a NS kick, a PWN if exists, instabilities, and the hot ejecta that expands to all directions, where heating is due to the reverse shock, nickel decay (nickel bubbles), and initial heat from the explosion (see \cite{SokerShishkin2025Vela} for a discussion of these processes). Another process is launching two unequal jets in some jet-launching episodes, which results from the short duration of the jet-launching episode. Namely, the intermittent accretion disk exists for a time shorter, or not much longer, than the dynamical time of the accretion disk, and hence, the accretion disk has no time to fully relax (see discussion in \citealt{Soker2024N63A}).

On panel b of Figure \ref{fig:SNRG0901}, I connect Rim 4N and Rim 4S with a double-sided double-lined red arrow, and so the pair of the faint zones that I term Valley N and Valley S. The centers of the two arrows cross at the same point on the two previously identified symmetry axes (dashed pale-blue and dotted-greed double-sided arrows) that compose the main-jet axis of SNR~G0.9+0.1 (Section \ref{subsec:NorthernEar}); the crossing point at the center of the double-lined red arrow is not at the center of the dashed and dotted lines of the main-jet axis. I also identify Rim 3N and Rim 3S as a pair. The four double-sided arrows on panel b of Figure \ref{fig:SNRG0901} form the `wind-rose' of the point-symmetric SNR~G0.9+0.1.  

On panel b of Figure \ref{fig:SNRG0901}, I mark Valley S, several filaments to the west of the main-jet axis, and the western boundary of the bright shell with curved dashed pale-blue lines. In panel d of Figure \ref{fig:SNRG0901}, I rotate this `western dashed pale blue lines structure' by $180^\circ$ around the cross point of the double-sided arrows and find that it matches some features on the east side of the main-jet axis, in particular Rim 3S and Rim 4S. Valley S does not fully overlap with Valley N. In panel c of Figure \ref{fig:SNRG0901}, I displaced the rotated western dashed pale blue lines structure to the southeast to make the rotated Valley S to overlap Valley N. 

With the four double-sided arrows and the $180^\circ$ rotation of the western dashed pale blue lines structure, I demonstrated the point-symmetric morphology of SNT~G0.9+0.1. Considering the smearing processes mentioned above, I think identifying the point-symmetric morphology is robust. I attribute it to the operation of jets. Simulations show that a pair of jets can form more than one pair of morphological features \citep{Braudoetal2025}, while some pairs of jets that participate in the explosion process are choked deep in the core and leave no clear imprints at the CCSNR phase. Therefore, it is not straightforward to determine the number of jet pairs that shaped the SNR~G0.9+0.1.  

\section{Random filaments cannot indicate the explosion mechanism}
\label{sec:Instabilities}

In this section, I argue that filaments and holes, as simulated by \cite{Orlandoetal2025Filaments} and \cite{Orlandoetal2025Holes} to explain some filaments and holes in the morphology of Cassiopeia A, cannot distinguish between the two explosion mechanisms, i.e., cannot resolve the question on whether the neutrino-driven mechanism or the JJEM exploded Cassiopeia A. If the filaments have a point-symmetric morphology, as \cite{BearSoker2025} identified in Cassiopeia A, these filaments imply the JJEM; \cite{Orlandoetal2025Filaments} consider different filaments, those in the inner zone of the CCSNR. 

\begin{figure*}
\begin{center}
\includegraphics[trim=0.0cm 11.0cm 0.0cm 0.5cm,scale=0.65]{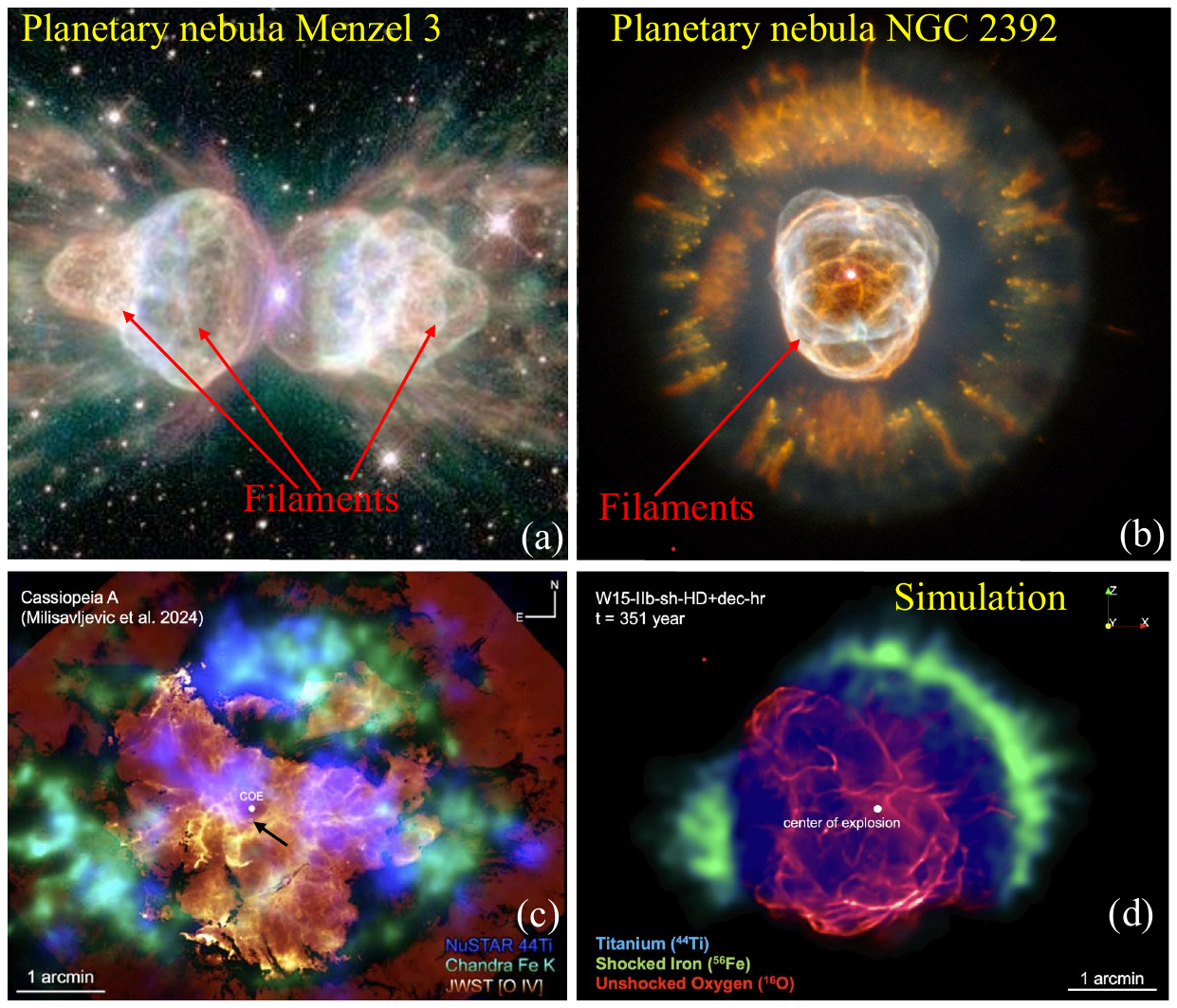} 
\caption{Astronomical objects with filamentary structures.  
(a) The inner part of the planetary nebula Menzel 3 (the `Ant Nebula'; Credit: NASA, ESA and the Hubble Heritage Team, STScI/AURA).
(b) The planetary nebula NGC 2392
[Credit: NASA, Andrew Fruchter and the ERO Team: Sylvia Baggett, (STScI), Richard Hook (ST-ECF), Zoltan Levay (STScI)]. It has an inner filamentary zone.  
The lower panels are adapted from \cite{Orlandoetal2025Filaments}. (c) An image of the center of Cassiopeia A that \cite{Orlandoetal2025Filaments} adapted from \cite{Milisavljevicetal2024}. Blue represents NuSTAR $^{44}$Ti emission, green represents Chandra Fe-K emission, and red represents the unshocked ejecta from JWST observations. The small black circle that a black arrow points at is the location of the NS. (d) As the inset indicates, the numerical simulation results by \cite{Orlandoetal2025Filaments} show 3D volumetric renderings of the distributions of three elements. Although the simulations are based on the results of a simulated neutrino-driven explosion, the filamentary structures of the planetary nebulae imply such structures are not unique to the neutrino-driven explosion mechanism.  
}
\label{Fig:PNe}
\end{center}
\end{figure*}

In Figure \ref{Fig:PNe}, I present two planetary nebulae with filamentary structures, an image of the central region of SNR Cassiopeia A, and a result of a simulation by \cite{Orlandoetal2025Filaments}.  \cite{Orlandoetal2025Filaments} simulation is a continuation of a simulation of an explosion by the neutrino-driven explosion mechanism. \cite{Orlandoetal2025Filaments} claim their simulations produce the general filamentary structures observed in Cassiopeia A. The filamentary structure they obtain is qualitatively similar to the filamentary structure of the SNR S147, the “Spaghetti Nebula”, which has a prominent H$\alpha$ filamentary texture  (e.g., \citealt{Khabibullinetal2024} for a recent observational study of S147). My claim is that the formation of filamentary structures is not unique to the neutrino-driven explosion mechanism. Instabilities form such structures, which are also obtained in JJEM simulations \citep{Braudoetal2025} and observed in planetary nebulae. The upper panels of Figure \ref{Fig:PNe} show two such planetary nebulae, the inner part of Menzel 3 which is filamentary (e.g., \citealt{Guerreroetal2004M3}) and by its structure must have been shaped by jets (e.g., \citealt{AkashiSoker2008}), and NGC 2392, with a rich set of filaments in the inner region (e.g., \citealt{Guerreroetal2021Es}), and has been shaped by observed jets (e.g.,  \citealt{Giesekingetal1985, Guerreroetal2021Es}). Jets shaped tens more filamentary planetary nebulae.  
These images and simulations show that filaments are formed by instabilities that might occur in expanding nebulae/ejecta that interact with slower material and do not require a neutrino-driven explosion or the production of radioactive elements. Therefore, filamentary structures cannot teach us about the explosion mechanism of CCSNe.    

\begin{figure*}
\begin{center}
\includegraphics[trim=0cm 9.5cm 0.0cm 0cm,scale=0.65]{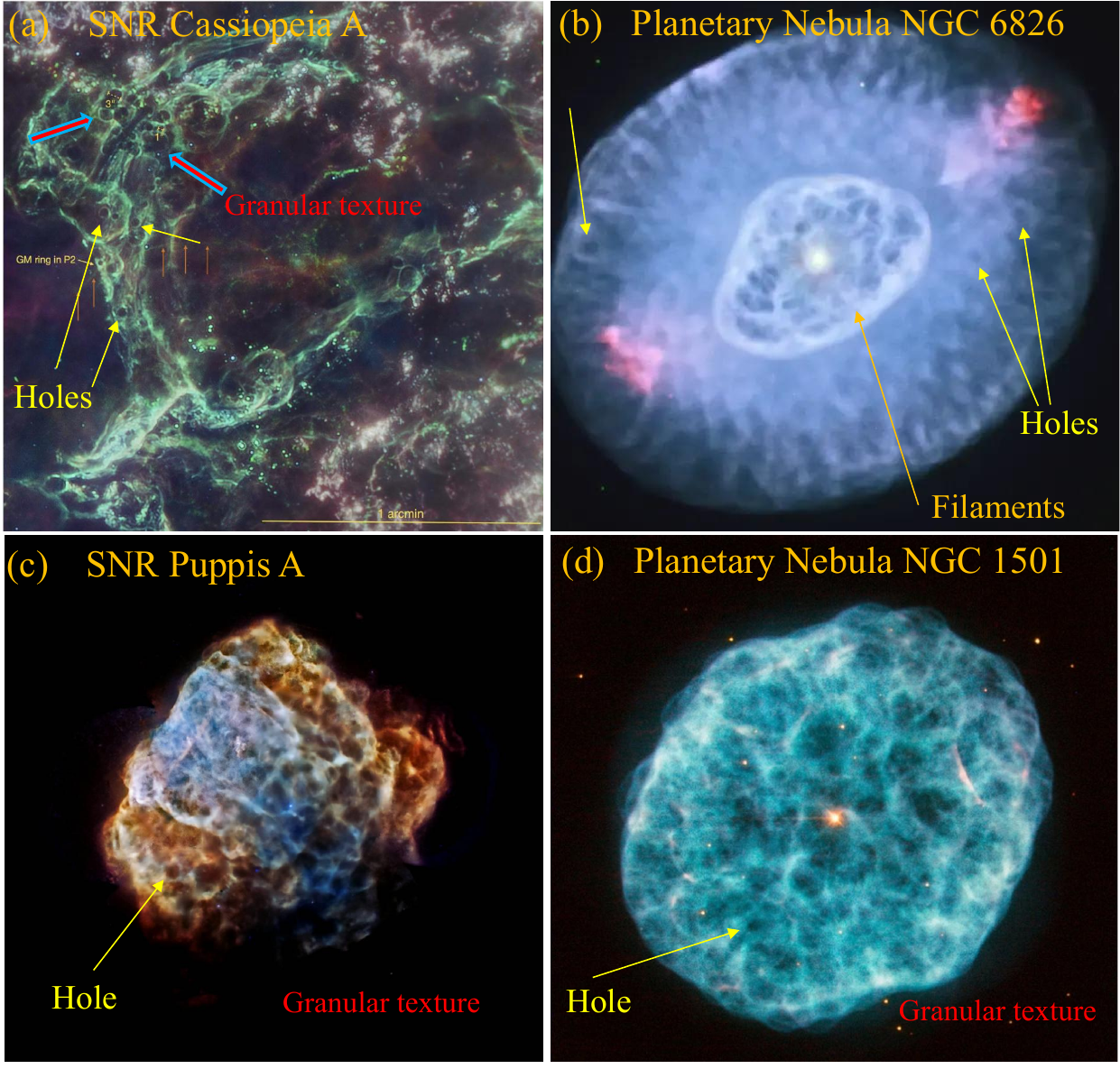} 
\caption{
Comparing holes in the `Green Monster' of SNR Cassiopeia A with a planetary nebula. 
(a) A JWST/NIRCam and MIRI image of Cassiopeia A adapted from \cite{DeLoozeetal2024}  (see also \citealt{Milisavljevicetal2024}). Green represents emission mostly in the MIRI/F1280W filter with contributions from MIRI/F1130W. I added arrows to point at a few holes out of the many that \cite{Orlandoetal2025Holes} studied with their simulations. 
(b)  An HST image of NGC 6826 (Credit: NASA / ESA / Judy Schmidt). Yellow arrows point at several holes. The inner zone is filamentary, and I point at three holes in the outer region (there are more holes).
(c) An X-ray image of SNR Puppis A with its granular texture (Credit: NASA/CXC/IAFE/; \citealt{Dubneretal2013}).
(d) A planetary nebula, NGC 1501, with its granular texture; an image adapted from ESA (Credit: ESA/Hubble \& NASA; acknowledgment: M. Canale).
The colors red, green, and blue represent the [N \textsc{II}], H$\alpha$, and [O \textsc{III}] emission (e.g., \citealt{Rubioetal2022}). It has a granular texture. 
}
\label{Fig:Holes}
\end{center}
\end{figure*}

The same conclusion holds for the holes in SNR Cassiopeia A. In panel a of Figure \ref{Fig:Holes}, I present the holes in the `green monster' of Cassiopeia A. This structure results from the interaction of the ejecta with circumstellar material. \cite{Orlandoetal2025Holes} simulated the formation of such holes. The formation of holes is not unique to the neutrino-driven mechanism, as, for example, holes also exist in some planetary nebulae (where neutrino and radioactive products play no role at all). On the image of the planetary nebula NGC 6826 
(e.g., \citealt{BarriaKimeswenger2018}), in panel b of Figure \ref{Fig:Holes}, I mark three holes, out of many more. Several holes that share boundaries form a granular texture. I point at two such areas in the `Green Monster' of Cassiopeia A with double-line arrows in panel a of Figure \ref{Fig:Holes}. The X-ray image of the SNR Puppis A, panel c of Figure \ref{Fig:Holes}, presents a granular texture on most of this SNR. \cite{Bearetal2025Puppis} noted the similar granular texture of the SNR Puppis A and those of some planetary nebulae, e.g., A72. \cite{Bearetal2025Puppis} argued that the morphological similarity suggests that the granular texture results from instabilities and cannot be attributed to a neutrino-driven explosion or to radioactive products (nickel bubbles). Instabilities can develop in the interaction of different parts of the ejecta or the interaction of the main ejecta with a previously ejected mass. In panel d of Figure \ref{Fig:Holes}, I present an image of the planetary nebula NGC 1501 (e.g., \citealt{Sabbadinetal2000, Rubioetal2022}) that has a granular texture.    

\section{Summary}
\label{sec:Summary}

This study is another step towards finding the explosion mechanism of CCSNe. I identified a point-symmetrical morphology in the SNR~G0.9+0.1, which the JJEM predicts and can explain, while the neutrino-driven mechanism does not. 

The most prominent structure outside the bright PWN is the ear to the northwest, Ear N, with its two rims, Rim 1 and Rim 2 (Figure \ref{fig:SNRG0901}). Figure \ref{Fig:KjPn8} presents an image of a jet-shaped planetary nebula that has rims on one lobe, and  Figure \ref{Fig:Hercules} presents the cluster of galaxies Hercules and its jets, with several jet-shaped rims on one side. In Section \ref{subsec:NorthernEar}, I considered the similar one-sided rim structure in these three objects to argue that Ear N of SNR~G0.9+0.1 was also shaped by at least two jets.    

In Section \ref{subsec:NorthernEar}, I further propose that the bright region south of the main shell of SNR~G0.9+0.1 is a jet-shaped blowout, i.e., a jet that broke out from the main SNR shell inflated this region. I based this on the similarity of the blowout of SNR~G0.9+0.1 with that of SNR~G309.2-00.6 that Figure \ref{fig:SNRG309} presents. The jet that inflated the blowout, I proposed there, is the jet's counter jet that shaped Rim 1, which is the outer rim of Ear N. I found two symmetry axes, the dashed-pale-blue and dotted green double-sided arrows on panel b of Figure \ref{fig:SNRG0901}, with only $6^\circ$ between them. Together, I took these two symmetry axes to define the main-jet axis of SNR~G0.9+0.1 (or, specifically, between them). In Section \ref{subsec:point}, I added two symmetry axes, the double-sided double-lined red arrows on panel b of Figure \ref{fig:SNRG0901}; the four double-sided arrows form the `wind-rose' of the point-symmetric morphology of SNR~G0.9+0.1. 

A point symmetric morphology of a SNR~G0.9+0.1 implies the JJEM and almost rules out the neutrino-driven mechanism. Some other morphological features cannot distinguish between the two explosion mechanisms.  In Section \ref{sec:Instabilities}, I presented images demonstrating that filaments, holes, and granular textures are common to some planetary nebulae and some CCSNe. These structures result from instabilities that exist in the neutrino-driven mechanism, as \cite{Orlandoetal2025Filaments} and \cite{Orlandoetal2025Holes} simulated to reproduce some morphological features of Cassiopeia A, they exist in the JJEM \citep{Braudoetal2025}, and planetary nebulae. Although the filaments in the central regions of Cassiopeia A that result from instabilities cannot distinguish between the two explosion mechanisms, some filaments on the outskirts that have opposite counterparts are part of the point-symmetric morphology of Cassiopeia A and, hence, support the JJEM \citep{BearSoker2025}.

\section*{Acknowledgements}

I thank Dmitry Shishkin and Ealeal Bear for their helpful comments and discussions. An Asher Space Research Institute grant at the Technion and the Charles Wolfson Academic Chair at the Technion supported this research. 


\end{document}